\documentclass[12pt]{article}
\usepackage{amsfonts}
\usepackage{epsfig,amssymb,euscript}
\usepackage{amsmath,amscd}

\numberwithin{equation}{section}
\newcommand{\be}{\begin{eqnarray}}
\newcommand{\ee}{\end{eqnarray}}
\newcommand{\bea}{\begin{eqnarray}}
\newcommand{\eea}{\end{eqnarray}}
\newcommand{\ba}{\begin{array}}
\newcommand{\ea}{\end{array}}
\newcommand{\nn}{\nonumber \\}

\begin{document}



\begin{titlepage}

\vfill

\begin{flushright}
hep-th/0610128\\
\end{flushright}

\vfill

\begin{center}

\baselineskip=16pt

{\Large\bf Maximally Minimal Preons in Four Dimensions}

\vskip 1.3cm

Jai Grover\footnote{Email: jg372@cam.ac.uk} and Jan B.
Gutowski\footnote{Email: J.B.Gutowski@damtp.cam.ac.uk}

\vskip 1cm

{\small{\it
DAMTP, Centre for Mathematical Sciences\\
University of Cambridge\\
Wilberforce Road, Cambridge, CB3 0WA, UK}}

\vskip 2cm

Wafic Sabra\footnote{Email:  ws00@aub.edu.lb}

\vskip .6cm {\small{\it
Centre for Advanced Mathematical Sciences and Physics Department\\
American University of Beirut\\ Lebanon}}

\end{center}

\begin{center}
\textbf{Abstract}
\end{center}

\begin{quote}

Killing spinors of $N=2$, $D=4$ supergravity are examined using the
spinorial geometry method, in which spinors are written as
differential forms. By making use of methods developed in
\cite{papadgran2006b} to analyze preons in type IIB
supergravity, we show that there are no simply connected
solutions preserving exactly 3/4 of the supersymmetry.

\end{quote}

\vfill

\end{titlepage}

\section{Introduction}

The classification of supersymmetric solutions has been an active area of
research due to the importance of these solutions in string and M-theory.
Many years ago, Tod was able to find all metrics admitting supercovariantly
constant spinors in $N=2$, $D=4$ ungauged minimal supergravity \cite{tod}.
In recent years and motivated by the work of \cite{tod} progress has been
made in the classification of supersymmetric solutions and in particular for
lower dimensional gauged supergravity theories \cite{classify,gutsab2006,
klemm2003,klemm2004}. The basic idea in this classification is to construct
differential forms as bilinears from the supercovariantly constant spinor.
The algebraic and differential equations satisfied by these forms are then
used to deduce the metric and the bosonic fields of the supergravity theory.

In our present work we will focus on the classification of supersymmetric
solutions of $N=2$, $D=4$ gauged supergravity. In the light of the the
AdS/CFT correspondence \cite{maldacena}, these solutions can shed light on
CFT in three dimensions. From the CFT\ point of view, solutions preserving
fractions of supersymmetry can be regarded as an expansion of the theory
around non-zero vacuum expectation values of certain operators. Moreover,
the classification of supersymmetric solutions is also relevant for the
construction of microstates for supersymmetric black holes \cite{micro}.

The classification of lightlike and timelike solutions preserving fractions
of supersymmetry of minimal gauged supergravity in four dimensions has been
performed in \cite{klemm2003, klemm2004}. In particular, in \cite{klemm2004}
it was shown that a configuration which admits a null Killing spinor can be
either 1/4 or 1/2 but not exactly 3/4 supersymmetric.

In this paper, we will show that, as in the five dimensional case
\cite{gutsab2006}, there are no simply connected
3/4 supersymmetric solutions in the theory of
$N=2 $, $D=4$ gauged supergravity irrespective of the nature of the
solutions. In our analysis, it will be particularly useful to consider the
spinors as differential forms \cite{lawson, wang, harvey}. This method of
writing spinors as forms has been used to classify solutions of supergravity
theories in ten and eleven dimensions (see for example \cite{papadgran2006b,
papadgran2005a,papadgran2005b, papadgran2006a}.)

The plan of the paper is as follows. In Section 2, the theory of $N=2$, $D=4$
gauged minimal supergravity is presented and it is shown how spinors of the
theory can be written as differential forms. The gauge freedom present in
the theory is used to reduce a spinor to one of three ``canonical'' forms. A
$Spin(3,1)$-invariant non-degenerate bilinear form $B$ on the space of
spinors is also defined. In Section 3, it is shown that solutions preserving
$3/4$ of the supersymmetry can be placed into three classes according as to
the canonical form of the spinor which is orthogonal (with respect to $B$)
to the Killing spinors. This method of characterizing supersymmetric
solutions by the spinors which are orthogonal to the Killing spinors was
originally developed in \cite{papadgran2006b} where it was used to show that
there are no preons in type IIB supergravity. The integrability conditions
of the Killing spinor equations, for all three possible types of solution,
fix the gauge field strengths to vanish, and constrain the spacetime
geometry to be locally isometric to $AdS_{4}$.
Furthermore, $AdS_4$ is a maximally supersymmetric solution
of $N=2$, $D=4$ minimal gauged supergravity.
It therefore follows that there can be no simply connected
\textit{exactly} $3/4$ supersymmetric solutions of $N=2$, $D=4$
minimal gauged supergravity.

\section{Supersymmetric solutions of $N=2$ supergravity}

\subsection{Minimal $N = 2$ gauged supergravity}

In this section we summarize some of the properties of the minimal gauged
$N=2$ supergravity theory in four dimensions, and also describe how to write
Killing spinors of this theory as differential forms. The bosonic action of
minimal $N=2$, $D=4$ gauged supergravity is \cite{freedman1977},
\cite{fradkin1976}

\begin{equation}
S=\int d^{4}x \sqrt{-g} \big({\frac{1}{4}}R-{\frac{1}{4}} F_{\mu
\nu}F^{\mu\nu} +{\frac{3}{2\ell^{2}}}\big)\ ,
\end{equation}

where $\ell$ is a nonzero real constant. The metric has signature $(-,+,+,+)$.
We shall consider solutions which preserve some proportion of the
supersymmetry; hence there are Killing spinors $\epsilon $ satisfying the
Killing spinor equation

\begin{eqnarray}  \label{killsp1}
D_{\mu} \epsilon = \nabla_{\mu} \epsilon + {\frac{1}{2\ell}}\gamma_{\mu}
\epsilon + {\frac{i}{4}}F_{\nu_1 \nu_2}\gamma^{\nu_1 \nu_2}\gamma_{\mu}
\epsilon -{\frac{i }{\ell}} A_\mu \epsilon =0
\end{eqnarray}

where

\begin{equation}
\nabla _{\mu}=\partial_{\mu}+{\frac{1}{4}} \omega _{\mu,\nu_{1}\nu_{2}}
\gamma^{\nu_{1}\nu_{2}}
\end{equation}
\ \ \ \ \

and $F=dA$ is the $U(1)$ gauge field strength. Maximally supersymmetric
solutions of this theory have $F=0$ and are locally isometric to $AdS_4$.
More generally,
supersymmetric solutions must preserve either 2, 4, 6 or 8 of the
supersymmetries, because the Killing spinor equation ({\ref{killsp1}}) is
linear over $\mathbb{C}$. Hence, {\textit{preonic}} solutions which for this
theory would preserve exactly $7/8$ of the supersymmetry are not possible.
Preons have been examined in ten and eleven-dimensional supergravity
theories \cite{papadgran2006b,bandos2001, duff2003, bandos2003, bandos2006,
hull2003}. In addition, many examples of $1/4$ and $1/2$-supersymmetric
solutions of $N=2, D=4$ gauged supergravity are known \cite{klemm2003,
klemm2004,chamsab2000, sabra1999}. It is therefore natural to examine
whether it is possible to construct solutions preserving exactly $3/4$ of
the supersymmetry.

\subsection{Spinors in four dimensions}

The spinors $\epsilon$ appearing in the Killing spinor equation
({\ref{killsp1}}) are Dirac spinors. Following \cite{lawson, wang, harvey} these
spinors can be written as complexified forms on $\mathbb{R}^2$; if $\Delta$
denotes the space of Dirac spinors then
$\Delta = \Lambda^* (\mathbb{R}^2)\otimes \mathbb{C}$.
A generic spinor $\eta$ can therefore be written as

\begin{equation}
\eta =\lambda 1+\mu ^{i}e^{i}+\sigma e^{12}
\end{equation}

where $e^{1}$, $e^{2}$ are 1-forms on $\mathbb{R}^{2}$, and $i=1,2$;
 $e^{12}=e^{1}\wedge e^{2}$. $\lambda $, $\mu^{i}$ and $\sigma $
 are complex functions.

The action of $\gamma$-matrices on these forms is given by

\begin{eqnarray}
\gamma_0 &=& - e^2 \wedge + i_{e^2}  \nn
\gamma_1 &=& e^1 \wedge +i_{e^1}  \nn
\gamma_2 &=& e^2 \wedge + i_{e^2}  \nn
\gamma_3 &=& i( e^1 \wedge - i_{e^1}) \ .
\end{eqnarray}

$\gamma_5$ is defined by

\begin{eqnarray}
\gamma_5 = i \gamma_{0123}
\end{eqnarray}

and satisfies

\begin{eqnarray}
\gamma_5 1 = 1 , \quad \gamma_5 e^{12} = e^{12} , \quad \gamma_5 e^i = -e^i
\ \ i=1, 2 \ .
\end{eqnarray}

The charge conjugation operator $C$ is defined by

\begin{eqnarray}
C 1 = -e^{12}, \quad C e^{12} = 1 \quad C e^i = - \epsilon_{ij} e^j \ \ i=1,2
\end{eqnarray}

where $\epsilon_{ij}=\epsilon^{ij}$ is antisymmetric with $\epsilon_{12}=1$.
We also use the convention $\epsilon_{0123}=1$.

We note the useful identities

\begin{eqnarray}  \label{eqn:crelat}
(\gamma_\mu)^* = \gamma_0 C \gamma_\mu \gamma_0 C
\end{eqnarray}

and

\begin{eqnarray}
C \gamma_m^{*} &=& \gamma_m C  \nn
C \gamma_0^{*} &=& -\gamma_0 C
\end{eqnarray}

and

\begin{eqnarray}
(\gamma_0)_{ab} = - (\gamma_0^*)_{ba}, \quad (\gamma_m)_{ab} =
(\gamma_m^*)_{ba}
\end{eqnarray}

for $m = 1, 2, 3$; where $(\gamma_\mu)_{ab} \equiv \delta_{ac}
(\gamma_\mu)^c{}_b $.

It will be particularly useful to complexify the gamma-operators via

\begin{eqnarray}
\gamma_{+} &=& {\frac{1 }{\sqrt{2}}} (\gamma_2 + \gamma_0) = \sqrt{2} i_{e^2}
\nn
\gamma_{-} &=& {\frac{1 }{\sqrt{2}}} (\gamma_2 - \gamma_0) = \sqrt{2} e^2
\wedge  \nn
\gamma_{1} &=& {\frac{1 }{\sqrt{2}}} (\gamma_1 + i \gamma_3) = \sqrt{2}
i_{e^1}  \nn
\gamma_{\bar{1}} &=& {\frac{1 }{\sqrt{2}}} (\gamma_1 - i \gamma_3) = \sqrt{2}
e^1 \wedge
\end{eqnarray}

where the metric components in the null basis are given by $g_{+-} = 1,
g_{1\bar{1}} = 1$.

\subsection{Gauge transformations and canonical spinors}

There are two types of gauge transformation which can be used to simplify
the Killing spinors of this theory. First, there are local $U(1)$ gauge
transformations of the type

\begin{eqnarray}  \label{u1gauge}
\epsilon \rightarrow e^{i \theta} \epsilon
\end{eqnarray}

for real functions $\theta$, and there are also local $Spin(3,1)$ gauge
transformations of the form

\begin{eqnarray}
\epsilon \rightarrow e^{{\frac{1 }{2}} f^{\mu \nu} \gamma_{\mu \nu}} \epsilon
\end{eqnarray}

for real functions $f^{\mu \nu}$.

Note in particular that $\gamma_{12}, \gamma_{13}, \gamma_{23}$
generate $SU(2)$ transformations which act (simultaneously)
on both $1, e^{12}$ and $e^1$, $e^2$. In particular, $\gamma_{13}$ acts via
\begin{eqnarray}
1 \rightarrow e^{i \theta} 1, \quad e^1 \rightarrow e^{-i \theta} e^1, \quad
e^2 \rightarrow e^{i \theta} e^2, \quad e^{12} \rightarrow e^{-i \theta}
e^{12}
\end{eqnarray}
for $\theta \in \mathbb{R}$. Furthermore, $\gamma_{02}$ generates a scaling
of the form
\begin{eqnarray}
1 \rightarrow e^{x} 1, \quad e^1 \rightarrow e^{x} e^1, \quad e^2
\rightarrow e^{-x} e^2, \quad e^{12} \rightarrow e^{-x} e^{12}
\end{eqnarray}
for $x \in \mathbb{R}$.

Applying the $SU(2)$ transformation on a general spinor of the form

\begin{eqnarray}
\epsilon = \lambda 1 + \mu^p e^p + \sigma e^{12}
\end{eqnarray}

allows us to set $\sigma=0$ and $\lambda \in \mathbb{R}$ so that

\begin{eqnarray}
\epsilon = \lambda 1 + \mu^1 e^1 + \mu^2 e^2 \ .
\end{eqnarray}

There are then various cases to consider.

First, suppose that $\mu^2 \neq 0$. Then consider the $Spin(3,1)$ gauge
transformation generated by $\gamma_{01}-\gamma_{12}$
and $\gamma_{03}+\gamma_{23}$, which acts via

\begin{eqnarray}
1 \rightarrow 1, \quad e^1 \rightarrow e^1, \quad e^2 \rightarrow
-2(y+ix)e^1 + e^2, \quad e^{12}\rightarrow 2(y-ix)1 + e^{12}
\end{eqnarray}

where $x,y \in \mathbb{R}$ are two gauge parameters. This transformation can
be used to set $\mu^1 = 0$, leaving

\begin{eqnarray}
\epsilon = \lambda 1 + \mu^2 e^2 \ .
\end{eqnarray}

If $\lambda \neq 0$, we can use the scaling generated by $\gamma_{02}$ to
obtain

\begin{eqnarray}  \label{can1}
\epsilon = 1 + \mu^2 e^2 \ .
\end{eqnarray}

However, if $\lambda=0$, then by combining the scaling
generated by $\gamma_{02}$ with a $SU(2)$ transformation
generated by $\gamma_{13}$ we can take

\begin{eqnarray}  \label{can2}
\epsilon = e^2 \ .
\end{eqnarray}

If instead $\mu^2 = 0$, then there again two cases. If $\lambda \neq 0$,
then by combining the scaling generated by $\gamma_{02}$ with a $SU(2)$
transformation generated by $\gamma_{13}$ we can set

\begin{eqnarray}  \label{can3}
\epsilon = 1 + \mu^1 e^1
\end{eqnarray}
where $\mu^1 \in \mathbb{C}$.

If however $\lambda =0$, then by using a $SU(2)$ transformation together
with the scaling generated by $\gamma_{02}$, the spinor can be
written as ({\ref{can2}}).

So, one can always use $Spin(3,1)$ gauge transformations to write a
\textit{single} spinor as

\begin{eqnarray}  \label{eqn:spin2}
\epsilon = e^2
\end{eqnarray}

or

\begin{eqnarray}  \label{eqn:spin3}
\epsilon = 1+ \alpha e^1
\end{eqnarray}

or

\begin{eqnarray}  \label{eqn:spin5}
\epsilon = 1+\beta e^2
\end{eqnarray}

for some functions $\alpha, \beta \in \mathbb{C}$.

\subsection{A $Spin(3,1)$ invariant inner product on spinors}

In order to analyze the $3/4$ supersymmetric solutions it is necessary to
construct a non-degenerate inner product on the space of spinors. We first
define a Hermitian inner product on the space of spinors via

\begin{equation}
\langle
z^{0}1+z^{1}e^{1}+z^{2}e^{2}+z^{3}e^{12},w^{0}1+w^{1}e^{1}+w^{2}e^{2}+w^{3}e^{12}
\rangle ={\bar{z}}^{q}w^{q}
\end{equation}

summing over $q=0,1,2,3$. However, $\langle ,\rangle $ is not $Spin(3,1)$
gauge-invariant. To rectify this, we define an inner product $B$ by

\begin{eqnarray}
B(\eta, \epsilon) = \langle C \eta^{*} , \epsilon \rangle
\end{eqnarray}

then it is straightforward to show that $B$ satisfies

\begin{eqnarray}
B(\eta, \epsilon)+B(\epsilon,\eta)&=&0 \cr B(\gamma_\mu
\eta,\epsilon)-B(\eta, \gamma_\mu \epsilon) &=&0 \cr B(\gamma_{\mu \nu}
\eta, \epsilon)+ B(\eta, \gamma_{\mu \nu} \epsilon) &=&0
\end{eqnarray}
for all spinors $\eta, \epsilon$.

The last of the above constraints implies that $B$ is $Spin(3,1)$ invariant.
Note that $B$ is linear over $\mathbb{C}$ in both arguments. The inner
product $B$ is non-degenerate: if $B(\epsilon,\eta)=0$ for all $\eta$
then $\epsilon=0$.

To show the $Spin(3,1)$ invariance of $B$ we consider

\begin{eqnarray}
B(\gamma_{\mu \nu} \eta, \epsilon) &=& \langle C\gamma_{\mu\nu}^{*}\eta^{*},
\epsilon \rangle \ .
\end{eqnarray}

Then for $m,n = 1,2,3$.

\begin{eqnarray}
B(\gamma_{mn} \eta, \epsilon) &=& \langle \gamma_{mn}C \eta^{*}, \epsilon
\rangle  \nn
&=& (\gamma_{mn})^{*}_{ab}(C\eta)^b \epsilon^a  \nn
&=& -(\gamma_{mn})_{ba}(C\eta)^b \epsilon^a  \nn
&=& -B(\eta, \gamma_{mn}\epsilon)
\end{eqnarray}

and

\begin{eqnarray}
B(\gamma_{0n} \eta, \epsilon) &=& \langle -\gamma_{0n}C \eta^{*}, \epsilon
\rangle  \nn
&=& -(\gamma_{0n})^{*}_{ab}(C\eta)^b \epsilon^a  \nn
&=& -(\gamma_{0n})_{ba}(C\eta)^b \epsilon^a  \nn
&=& -B(\eta, \gamma_{0n}\epsilon) \ .
\end{eqnarray}

We have then verified the $Spin(3,1)$ invariance of the product.

\section{$3/4$ Supersymmetric Solutions}

We now proceed to examine solutions preserving six out of the eight allowed
supersymmetries. This implies the existence of three Killing spinors, which
we shall denote by $\epsilon_0, \epsilon_1, \epsilon_2$, which are linearly
independent over $\mathbb{C}$. More precisely, it is assumed that there is
some open neighbourhood $U$ such that at every point in $U$, $\epsilon_0,
\epsilon_1, \epsilon_2$ are linearly independent over $\mathbb{C}$.

Suppose we denote the span (over $\mathbb{C}$) of $\epsilon_0$,
$\epsilon_1$, $\epsilon_2$ by $W$. Any complex three-dimensional
subspace of $\mathbb{C}^4$ can be uniquely specified by its one
(complex) dimensional orthogonal
complement with respect to the standard inner product on $\mathbb{C}^4$. It
follows that one can specify $W$ via its orthogonal complement with respect
to $B$. If the one dimensional $B$-orthogonal subspace to $W$ is
spanned by $\epsilon$, one has

\begin{eqnarray}
W = W_\epsilon = \{ \psi \in \Delta : B(\psi, \epsilon)=0 \}
\end{eqnarray}

for some fixed non-vanishing $\epsilon \in \Delta $. As $B$ is $Spin(3,1)$
invariant, it will be most convenient to use $Spin(3,1)$ gauge in order to
write the spinor $\epsilon $ in one of its canonical forms.

If $\epsilon=1+\alpha e^1$ then $W$ is spanned by $\eta_0=1$, $\eta_1=e^1$,
$\eta_2=e^2 - \alpha e^{12}$. If $\epsilon=1+ \beta e^2$ then $W$ is spanned
by $\eta_0= 1$, $\eta_1=e^2$, $\eta_2 = e^1 + \beta e^{12}$. If $\epsilon
=e^2$ then $W$ is spanned by $\eta_0=1$, $\eta_1=e^2$, $\eta_2= e^{12}$.

In all cases the Killing spinors $\epsilon_0$, $\epsilon_1$, $\epsilon_2$
are related to the spinors $\eta_A$ for $A=0,1,2$ via

\begin{eqnarray}
\epsilon_A = z_A{}^B \eta_B
\end{eqnarray}

where $z$ is a complex $3 \times 3$ matrix such that $\det z \neq 0$.

In order to analyze the solutions we shall consider the integrability
conditions associated with the killing spinor equations ({\ref{killsp1}).
These can be written as }

\begin{eqnarray}
\big[{\frac{1 }{\ell}}*F_{\mu \nu}\gamma_5 - i {\frac{1 }{\ell}} F_{\mu \nu}
-i \big({\frac{1}{\ell}}F^{\nu_1}{}_{[\mu}\gamma_{\nu]\nu_1}\big)
- i\big( \nabla_{[\mu}F_{\nu]}{}^{\nu_1}\gamma_{\nu_1}\big)  \nn
+ \big({\frac{1 }{2\ell^2}}\gamma_{\mu \nu} + {\frac{1}{4}}
R^{\nu_1\nu_2}{}_{\mu \nu} \gamma_{\nu_1\nu_2} + {\frac{1}{4}}F_{\nu_1\nu_2}
F^{\nu_1\nu_2}\gamma_{\mu \nu} -
F^{\nu_1\nu_2}F_{\nu_2[\mu}\gamma_{\nu]\nu_1}\big)  \nn
-i\big({\frac{1}{2}}\gamma_{\nu_1\nu_2[\mu}\nabla_{\nu]}
F^{\nu_1\nu_2} \big) \big] \epsilon_A = 0 \ .
\end{eqnarray}

for $A=0, 1, 2$. This constraint is equivalent to

\begin{eqnarray}
{\tilde{R}}_{\mu \nu} \eta_A =0
\end{eqnarray}

where

\begin{eqnarray}
{\tilde{R}}_{\mu \nu} \eta_A \equiv \big({\frac{1}{2}} (S^2{}_{\mu
\nu})^{\nu_1\nu_2} \gamma_{\nu_1\nu_2} + i{\frac{1}{2}} (T^2{}_{\mu
\nu})^{\nu_1 \nu_2}\gamma_{\nu_1\nu_2} + i(T^1{}_{\mu
\nu})^{\nu_1}\gamma_{\nu_1}  \nn
+ (V^1{}_{\mu \nu})^{\nu_1}\gamma_{\nu_1}\gamma_5 + (V^5{}_{\mu \nu})
\gamma_5 -{\frac{i}{\ell}}F_{\mu \nu}\big)\eta_A
\end{eqnarray}

for $A=0,1,2$, with

\begin{eqnarray}  \label{eqn:F}
(S^2{}_{\mu \nu})^{\nu_1\nu_2} &=& {\frac{1 }{2\ell^2}}
\delta_{\mu}{}^{[\nu_1} \delta_{\nu}{}^{\nu_2]}
+ {\frac{1}{4}} R^{\nu_1\nu_2}{}_{\mu \nu}
\nn
&+&{\frac{1}{4}}F_{\nu_3\nu_4}F^{\nu_3\nu_4}
\delta_{\mu}{}^{[\nu_1}\delta_{\nu}{}^{\nu_2]} -
F^{\nu_3[\nu_1}F_{\nu_3[\mu}\delta_{\nu]}{}^{\nu_2]}  \nn
(T^2{}_{\mu \nu})^{\nu_1\nu_2} &=& -{\frac{1}{\ell}}F^{[\nu_2}{}_{[\mu}
\delta_{\nu]}{}^{\nu_1]}  \nn
(T^1{}_{\mu \nu})^{\nu_1} &=& -\nabla_{[\mu} F_{\nu]}{}^{\nu_1}  \nn
(V^1{}_{\mu \nu})^{\nu_1} &=& {\frac{1}{2}} \epsilon_{\nu_2\nu_3
[\mu}{}^{\nu_1} \nabla_{\nu]} F^{\nu_2\nu_3}  \nn
(V^5{}_{\mu \nu}) &=& {\frac{1}{\ell}}*F_{\mu \nu} \ .
\end{eqnarray}

In all cases, we shall show that the integrability condition
${\tilde{R}}_{\mu \nu} \eta_A=0$ for $A=0,1,2$ can be
used to obtain constraints that
are sufficient to fix $F=0$, and so $T^1=T^2=V^1=V^5=0$. Furthermore, in all
cases, the integrability conditions then imply that $S^2=0$, or equivalently

\begin{eqnarray}
R_{\mu \nu \nu_1 \nu_2} = -{\frac{2 }{\ell^2}} g_{\mu [\nu_1} g_{\nu_2] \nu}
\ .
\end{eqnarray}

This implies that the spacetime geometry is locally isometric to $AdS_4$.
However, it is known that $AdS_4$ is a maximally supersymmetric solution
of this theory, and that all maximally supersymmetric solutions must be
locally isometric to $AdS_4$.
Hence there can be no simply connected solutions preserving \textit{exactly} $3/4$ of the
supersymmetry.

In the following analysis, we present the integrability constraints used to
prove this for all possible types of $3/4$ supersymmetric solutions,
according as whether the Killing spinors $\epsilon_A$ are orthogonal to
$1+\alpha e^1$ or $1 + \beta e^2$ or $e^2$.

In what follows it will be convenient to suppress the $\mu \nu$ indices in
the tensors $S^2, T^2, T^1, V^1,V^5$ and $F$.

\subsection{Minimal solutions with $B$-orthogonal spinors to
$1 + \protect \alpha e^1$}

The integrability constraints obtained by requiring that ${\tilde{R}}_{\mu
\nu} 1=0$ are

\begin{eqnarray}  \label{eqn:11}
(S^2)^{+-} + (S^2)^{1\bar{1}} + V^5 + i(T^2)^{+-} + i(T^2)^{1\bar{1}}
- {\frac{i}{\ell}}F = 0
\end{eqnarray}

\begin{eqnarray}  \label{eqn:12}
i(T^1)^{\bar{1}} + (V^1)^{\bar{1}} = 0
\end{eqnarray}

\begin{eqnarray}  \label{eqn:13}
i(T^1)^- + (V^1)^- = 0
\end{eqnarray}

\begin{eqnarray}  \label{eqn:14}
(S^2)^{-\bar{1}} + i(T^2)^{-\bar{1}} = 0
\end{eqnarray}

the integrability constraints obtained by requiring that ${\tilde{R}}_{\mu
\nu} e^1 =0$ are

\begin{eqnarray}  \label{eqn:21}
i (T^1)^1 - (V^1)^1 =0
\end{eqnarray}

\begin{eqnarray}  \label{eqn:22}
(S^2)^{+-} - (S^2)^{1\bar{1}} + i(T^2)^{+-} -i(T^2)^{1\bar{1}} - V^5
-{\frac{i}{\ell}}F =0
\end{eqnarray}

\begin{eqnarray}  \label{eqn:23}
(S^2)^{-1} + i(T^2)^{-1}=0
\end{eqnarray}

\begin{eqnarray}  \label{eqn:24}
-i(T^1)^- +(V^1)^- = 0
\end{eqnarray}

and the integrability constraints obtained by requiring that
${\tilde{R}}_{\mu \nu}( e^2- \alpha e^{12}) =0$ are

\begin{eqnarray}  \label{eqn:31}
-\sqrt{2}\alpha(S^2)^{+1} -i \sqrt{2}\alpha(T^2)^{+1} + i(T^1)^+ - (V^1)^+ =
0
\end{eqnarray}

\begin{eqnarray}  \label{eqn:32}
-\sqrt{2}(S^2)^{+\bar{1}} -\sqrt{2}i(T^2)^{+\bar{1}} + i\alpha(T^1)^+ +
\alpha(V^1)^+= 0
\end{eqnarray}

\begin{eqnarray}  \label{eqn:33}
-(S^2)^{+-} + (S^2)^{1\bar{1}} -i(T^2)^{+-} + i(T^2)^{1\bar{1}}  \nn
- i\sqrt{2}\alpha (T^1)^1 -\sqrt{2} \alpha (V^1)^1 -V^5 - {\frac{i}{\ell}}F
= 0
\end{eqnarray}

\begin{eqnarray}  \label{eqn:34}
\alpha (S^2)^{+-} + \alpha (S^2)^{1\bar{1}} + i\alpha (T^2)^{+-}
+i\alpha(T^2)^{1\bar{1}}  \nn
+ i\sqrt{2}(T^1)^{\bar{1}}-\sqrt{2}(V^1)^{\bar{1}} -\alpha(V^5) +
{\frac{i\alpha}{\ell}}F = 0 \ .
\end{eqnarray}

By taking the real and imaginary parts of ({\ref{eqn:11}}) we see that

\begin{eqnarray}
(S^2)^{+-} + V^5 +i(T^2)^{1\bar{1}} =0
\end{eqnarray}

\begin{eqnarray}
(S^2)^{1\bar{1}} + i(T^2)^{+-} -{\frac{i}{\ell}}F = 0
\end{eqnarray}

and in the same way ({\ref{eqn:22}}) yields

\begin{eqnarray}
(S^2)^{+-} - V^5 - i(T^2)^{1\bar{1}} =0
\end{eqnarray}

\begin{eqnarray}
-(S^2)^{1\bar{1}} + i(T^2)^{+-} -{\frac{i}{\ell}}F = 0 \ .
\end{eqnarray}

These equations imply that

\begin{eqnarray}
(S^2)^{+-} = (S^2)^{1\bar{1}} = 0
\end{eqnarray}

\begin{eqnarray}  \label{Fcond}
i(T^2)^{+-} -{\frac{i}{\ell}}F = 0 \ .
\end{eqnarray}

{}From equations ({\ref{eqn:14}}) and ({\ref{eqn:23}}) we see

\begin{eqnarray}
(S^2)^{-1} = (T^2)^{-1} = 0
\end{eqnarray}

and from ({\ref{eqn:13}}) and ({\ref{eqn:24}})

\begin{eqnarray}
(T^1)^{-} = (V^1)^{-} = 0 \ .
\end{eqnarray}

Note that, upon comparison with ({\ref{eqn:F}}), imposing ({\ref{Fcond}})
forces all components of $F$ to vanish. Hence
$F = V^1 = V^5 = T^1 = T^2 = 0$, and by the above constraints
it follows that $S^2=0$ also. This implies
that the spacetime geometry is locally isometric to $AdS_4$.

\subsection{Minimal solutions with $B$-orthogonal spinors to
$1 + \protect \beta e^2$}

The integrability constraints obtained by requiring that ${\tilde{R}}_{\mu
\nu} 1=0$ are given by

\begin{eqnarray}  \label{eqn:b11}
(S^2)^{+-} + (S^2)^{1\bar{1}} + V^5 + i(T^2)^{+-} + i(T^2)^{1\bar{1}}
- {\frac{i}{\ell}}F = 0
\end{eqnarray}

\begin{eqnarray}  \label{eqn:b12}
i(T^1)^{\bar{1}} + (V^1)^{\bar{1}} = 0
\end{eqnarray}

\begin{eqnarray}  \label{eqn:b13}
i(T^1)^- + (V^1)^- = 0
\end{eqnarray}

\begin{eqnarray}  \label{eqn:b14}
(S^2)^{-\bar{1}} + i(T^2)^{-\bar{1}} = 0
\end{eqnarray}

as before. The constraints that follow from ${\tilde{R}}_{\mu \nu} e^2 =0$
are

\begin{eqnarray}  \label{eqn:b21}
i(T^1)^+ - (V^1)^+ = 0
\end{eqnarray}

\begin{eqnarray}  \label{eqn:b22}
-(S^2)^{+\bar{1}} - i(T^2)^{+\bar{1}} = 0
\end{eqnarray}

\begin{eqnarray}  \label{eqn:b23}
-(S^2)^{+-} + (S^2)^{1\bar{1}} - i(T^2)^{+-} +i(T^2)^{1\bar{1}} - V^5
-{\frac{i}{\ell}}F =0
\end{eqnarray}

\begin{eqnarray}  \label{eqn:b24}
i(T^1)^{\bar{1}} - (V^1)^{\bar{1}} = 0 \ .
\end{eqnarray}

Lastly the integrability constraints arising from ${\tilde{R}}_{\mu \nu}
(e^1 +\beta e^{12}) =0$ are

\begin{eqnarray}  \label{eqn:b31}
\sqrt{2} \beta (S^2)^{+1} + \sqrt{2} i\beta (T^2)^{+1} + i(T^1)^1 -(V^1)^1 =
0
\end{eqnarray}

\begin{eqnarray}  \label{eqn:b32}
(S^2)^{+-} - (S^2)^{1\bar{1}} + i(T^2)^{+-} - i(T^2)^{1\bar{1}} - \sqrt{2}
i\beta(T^1)^+ - \sqrt{2}\beta(V^1)^+  \nn
- V^5 -{\frac{i}{\ell}}F = 0
\end{eqnarray}

\begin{eqnarray}  \label{eqn:b33}
\sqrt{2} (S^2)^{-1} +\sqrt{2} i(T^2)^{-1} + i\beta(T^1)^1 + \beta(V^1)^1 = 0
\end{eqnarray}

\begin{eqnarray}  \label{eqn:b34}
-\beta(S^2)^{+-} - \beta(S^2)^{1\bar{1}} - i\beta(T^2)^{+-}
- i\beta(T^2)^{1\bar{1}} - \sqrt{2}i(T^1)^- + \sqrt{2}(V^1)^-  \nn
+ \beta V^5 -{\frac{i\beta}{\ell}}F = 0 \ .
\end{eqnarray}

By taking the real and imaginary parts of ({\ref{eqn:b11}}) and
({\ref{eqn:b23}}) respectively, we see that

\begin{eqnarray}
(S^2)^{+-} + V^5 +i(T^2)^{1\bar{1}} =0
\end{eqnarray}

\begin{eqnarray}
(S^2)^{1\bar{1}} + i(T^2)^{+-} -{\frac{i}{\ell}}F = 0
\end{eqnarray}

and

\begin{eqnarray}
-(S^2)^{+-} - V^5 + i(T^2)^{1\bar{1}} =0
\end{eqnarray}

\begin{eqnarray}
(S^2)^{1\bar{1}} - i(T^2)^{+-} -{\frac{i}{\ell}}F = 0 \ .
\end{eqnarray}

These equations imply that

\begin{eqnarray}
(T^2)^{+-} = (T^2)^{1\bar{1}} = 0
\end{eqnarray}

\begin{eqnarray}
-(S^2)^{+-} - V^5 = 0
\end{eqnarray}

\begin{eqnarray}  \label{bFcond1}
(S^2)^{1\bar{1}} - {\frac{i}{\ell}}F = 0 \ .
\end{eqnarray}

Comparing ({\ref{eqn:b12}}) and ({\ref{eqn:b24}}) we find that

\begin{eqnarray}
(T^1)^{\bar{1}} = (V^1)^{\bar{1}} = 0
\end{eqnarray}

and from ({\ref{eqn:b13}}) and ({\ref{eqn:b21}})

\begin{eqnarray}
(T^1)^{-} = (V^1)^{-} = (T^1)^{+} = (V^1)^{+} = 0 \ .
\end{eqnarray}

Substituting these results into ({\ref{eqn:b32}}), we find

\begin{eqnarray}
(S^2)^{+-} - V^5 = 0
\end{eqnarray}

\begin{eqnarray}  \label{bFcond2}
-(S^2)^{1\bar{1}} - {\frac{i}{\ell}}F = 0 \ .
\end{eqnarray}

In this case, we note that imposing both ({\ref{bFcond1}}) and
({\ref{bFcond2}}) forces all components of $F$ to vanish.
Hence $F = V^1 = V^5=T^1 = T^2 = 0$, and by the above constraints
it follows that $S^2=0$ also.
This implies that the spacetime geometry is locally isometric to $AdS_4$.

\subsection{Minimal solutions with $B$-orthogonal spinors to $e^2$}

The integrability constraints obtained by requiring that ${\tilde{R}}_{\mu
\nu} 1=0$ are given by

\begin{eqnarray}  \label{eqn:c11}
(S^2)^{+-} + (S^2)^{1\bar{1}} + V^5 + i(T^2)^{+-} + i(T^2)^{1\bar{1}}
- {\frac{i}{\ell}}F = 0
\end{eqnarray}

\begin{eqnarray}  \label{eqn:c12}
i(T^1)^{\bar{1}} + (V^1)^{\bar{1}} = 0
\end{eqnarray}

\begin{eqnarray}  \label{eqn:c13}
i(T^1)^- + (V^1)^- = 0
\end{eqnarray}

\begin{eqnarray}  \label{eqn:c14}
(S^2)^{-\bar{1}} + i(T^2)^{-\bar{1}} = 0
\end{eqnarray}

as before. The constraints that follow from ${\tilde{R}}_{\mu \nu} e^2 =0$
are

\begin{eqnarray}  \label{eqn:c21}
i(T^1)^+ - (V^1)^+ = 0
\end{eqnarray}

\begin{eqnarray}  \label{eqn:c22}
-(S^2)^{+\bar{1}} - i(T^2)^{+\bar{1}} = 0
\end{eqnarray}

\begin{eqnarray}  \label{eqn:c23}
-(S^2)^{+-} + (S^2)^{1\bar{1}} - i(T^2)^{+-} +i(T^2)^{1\bar{1}} - V^5
-{\frac{i}{\ell}}F =0
\end{eqnarray}

\begin{eqnarray}  \label{eqn:c24}
i(T^1)^{\bar{1}} - (V^1)^{\bar{1}} = 0 \ .
\end{eqnarray}

Lastly the integrability constraints arising from ${\tilde{R}}_{\mu \nu}
e^{12}=0 $ are

\begin{eqnarray}  \label{eqn:c31}
(S^2)^{+1} + i(T^2)^{+1} = 0 \ .
\end{eqnarray}

\begin{eqnarray}  \label{eqn:c32}
i(T^1)^+ +(V^1)^+ = 0
\end{eqnarray}

\begin{eqnarray}  \label{eqn:c33}
i(T^1)^1 + (V^1)^1 = 0
\end{eqnarray}

\begin{eqnarray}  \label{eqn:c34}
-(S^2)^{+-} - (S^2)^{1\bar{1}} - i(T^2)^{+-} - i(T^2)^{1\bar{1}} + V^5
-{\frac{i}{\ell}}F = 0 \ .
\end{eqnarray}

By taking the real and imaginary parts of ({\ref{eqn:c11}}) and
({\ref{eqn:c23}}) respectively, we see that

\begin{eqnarray}
(S^2)^{+-} + V^5 +i(T^2)^{1\bar{1}} =0
\end{eqnarray}

\begin{eqnarray}
(S^2)^{1\bar{1}} + i(T^2)^{+-} -{\frac{i}{\ell}}F = 0
\end{eqnarray}

and

\begin{eqnarray}
-(S^2)^{+-} - V^5 + i(T^2)^{1\bar{1}} =0
\end{eqnarray}

\begin{eqnarray}
(S^2)^{1\bar{1}} - i(T^2)^{+-} -{\frac{i}{\ell}}F = 0 \ .
\end{eqnarray}

These equations imply that

\begin{eqnarray}
(T^2)^{+-} = (T^2)^{1\bar{1}} = 0
\end{eqnarray}

\begin{eqnarray}
-(S^2)^{+-} - V^5 = 0
\end{eqnarray}

\begin{eqnarray}  \label{cFcond1}
(S^2)^{1\bar{1}} - {\frac{i}{\ell}}F = 0 \ .
\end{eqnarray}

Comparing ({\ref{eqn:c12}}) and ({\ref{eqn:c24}}) we find that

\begin{eqnarray}
(T^1)^{\bar{1}} = (V^1)^{\bar{1}} = 0
\end{eqnarray}

and from ({\ref{eqn:c13}}) and ({\ref{eqn:c21}})

\begin{eqnarray}
(T^1)^{-} = (V^1)^{-} = (T^1)^{+} = (V^1)^{+} = 0 \ .
\end{eqnarray}

Substituting these results into ({\ref{eqn:c34}}), we find

\begin{eqnarray}
(S^2)^{+-} - V^5 = 0
\end{eqnarray}

\begin{eqnarray}  \label{cFcond2}
-(S^2)^{1\bar{1}} - {\frac{i}{\ell}}F = 0 \ .
\end{eqnarray}

In this case, we note that imposing both ({\ref{cFcond1}}) and
({\ref{cFcond2}}) forces all components of $F$ to vanish.
Hence $F = V^1 = V^5 =T^1 = T^2 = 0$, and by the above constraints
it follows that $S^2=0$ also.
This implies that the spacetime geometry is locally isometric to $AdS_4$.

\section{Conclusion}

In conclusion, we have completed the work of \cite{klemm2004} and studied
configurations preserving only $3/4$ of supersymmetry for the theory of $N=2$
$D=4$ minimal gauged supergravity. In our analysis we have employed the
method of writing spinors of the theory as differential forms. Using the
gauge symmetries of the spinors, one is able to place solutions preserving
$3/4$ of supersymmetry into three classes. Furthermore, using the
integrability conditions of the Killing spinor equations coming from the
vanishing of the gravitino supersymmetric variations, it was shown that the
gauge field strengths must vanish. This means that the spacetime geometry is
locally isometric to $AdS_4$. Hence solutions which preserve $3/4$ of
the supersymmetry are locally maximally supersymmetric. Therefore there
can be no simply connected {\it{exactly}} $3/4$ supersymmetric solutions.

One subtlety which remains to be addressed is whether there exist
non-simply connected solutions preserving $3/4$ of the supersymmetry for which
$F=0$, and the spacetime geometry is some quotient of $AdS_4$ by a discrete subgroup of
the symmetry group $Spin(3,2)$. For example, in the analysis of preons in $D=11$ supergravity,
it was proven in \cite{papadgran2007} that all solutions
preserving $31/32$ of the supersymmetry must be locally isometric to
maximally supersymmetric solutions, which proves that there can
be no simply connected preons in eleven dimensions. Then
in \cite{ofarrill2007} it was shown that no quotient of a
maximally supersymmetric solution by a discrete subgroup of
its symmetry group can preserve $31/32$ of the supersymmetry.
It would be interesting to see if $3/4$-supersymmetric quotients
of $AdS_4$ can be excluded using similar reasoning.

\section*{Acknowledgments}

Jan Gutowski thanks the organizers of the Eurostring Workshop II, Brno
during which part of this work was completed. Jai Grover thanks the
Cambridge Commonwealth Trusts for financial support. The work of Wafic Sabra
was supported in part by the National Science Foundation under grant number
PHY-0313416 and PHY-0601213.

\end{document}